\documentclass[,
    ,final            
    ,numberedheadings 
    ,cmfonts          
  ]
  {aipproc}

\layoutstyle{6x9}

\usepackage{psfrag}
\usepackage{graphicx}
\usepackage{amsmath}
\usepackage{amssymb}
\usepackage{bbm}

\newcommand{\be}{\begin{eqnarray}}
\newcommand{\ee}{\end{eqnarray}}

\newcommand{\E}{\mathrm{e}}
\newcommand{\I}{\mathrm{i}}

\begin{document}

\title{RG flow of the Polyakov-loop potential\\- first status report -}

\classification{64.60.Ak, 25.75.Nq, 11.15.-q}
\keywords      {renormalization group, quark deconfinement, gauge theories}

\author{J. Braun}{
  address={Institute for Theoretical Physics, University of
  Heidelberg, 69120 Heidelberg}
}

\author{H. Gies}{
  address={Institute for Theoretical Physics, University of
  Heidelberg, 69120 Heidelberg}
}

\author{H.-J. Pirner}{
  address={Institute for Theoretical Physics, University of
  Heidelberg, 69120 Heidelberg}
}
\begin{abstract}

We study SU(2) Yang-Mills theory at finite temperature in the
framework of the functional renormalization group. We concentrate on
the effective potential for the Polyakov loop which serves as an order
parameter for confinement. In this first status report, we focus on
the behaviour of the effective Polyakov-loop potential at high
temperatures. In addition to the standard perturbative result, our
findings provide information about the ``RG improved'' backreactions
of Polyakov-loop fluctuations on the potential. We demonstrate that
these fluctuations establish the convexity of the effective
potential. 

\end{abstract}

\maketitle


\section{Introduction}

An understanding of strongly interacting matter at finite temperature is a
prominent problem of contemporary physics that deserves to be analyzed with
great effort in view of the current and future experiments at heavy ion
colliders. Since the forces between quarks as elementary constituents are
governed by a non-Abelian gauge theory, already the understanding of gauge
boson dynamics is an important challenge. In this latter case of pure
gluodynamics, the expected transition to a deconfined phase can be studied
with the aid of the Polyakov loop \cite{Polyakov:1978vu},
being the order parameter for this transition:
\be
          \mathcal P (\vec{x})=\frac{1}{N}\mathrm{Tr} _{\text{F}}
          P\exp\Big(i\bar g\int_0^\beta A_0^a(\vec{x},t) t^a dt\Big).
\ee
Here $t^a$ are the generators of $SU(N)$ and $\beta$ denotes the
inverse temperature. The subscript $\text{F}$ alludes to the fundamental
representation. The negative
logarithm of the Polyakov-loop expectation value can be interpreted as
the free energy of a single static fundamental color source
\cite{svet86}. In this sense, an infinite free energy associated with
confinement is indicated as $\langle \mathcal P\rangle \to 0$, whereas
$\langle \mathcal P\rangle\neq 0$ signals deconfinement.

Moreover, $\langle \mathcal P\rangle$ measures whether center symmetry, a
discrete symmetry of Yang-Mills theory, is realized by the thermodynamic
ensemble \cite{svet86,'tHooft:1979ui}. Gauge transformations which differ at
Euclidean times $x_0=0$ and $x_0=\beta$ by a center element of the gauge group
change $\mathcal P$ by a phase $\E^{2\pi\I k/N}$, $k$ integer, but leave the
action and the functional integration measure invariant. This implies that a
center-symmetric ground state automatically ensures $\langle \mathcal
P\rangle=0$, whereas deconfinement $\langle \mathcal P\rangle\neq 0$ is
related to the breaking of this symmetry.

In fact, lattice simulations have not only collected strong evidence
for a second order phase transition in $SU(2)$ Yang-Mills theory
\cite{McLerran:1981pb,Engels:1981qx}, but reveal moreover that the
critical exponents agree with those of a 3D $Z(2)$ Ising model
\cite{Fortunato:2000vf}. The latter corresponds exactly to the
conjectured universality class obtained from the Polyakov-loop
criterion \cite{Svetitsky:1982gs}.

In recent years, an effective theory consisting of gauge-invariant powers of
the Polyakov loop has been developed based on mean-field arguments, see
\cite{Pisarski:2002ji} for an overview. Such considerations show a
good agreement with lattice data \cite{Fiore:2004}. Moreover, inverse
Monte-Carlo techniques have recently facilitated a precise lattice
determination of the Polyakov-loop effective action for SU(2)
\cite{Dittmann:2002ef}. 

\begin{figure}[]
{
\psfrag{V}[Bl][Bl][0.5]{$\beta ^4 V(\beta\bar{g}\phi)$}
\psfrag{phi}[Bl][Bl][0.5]{$\beta\bar{g}\phi$}
\hspace*{0.1cm}
\includegraphics[height=6.8cm,angle=270]{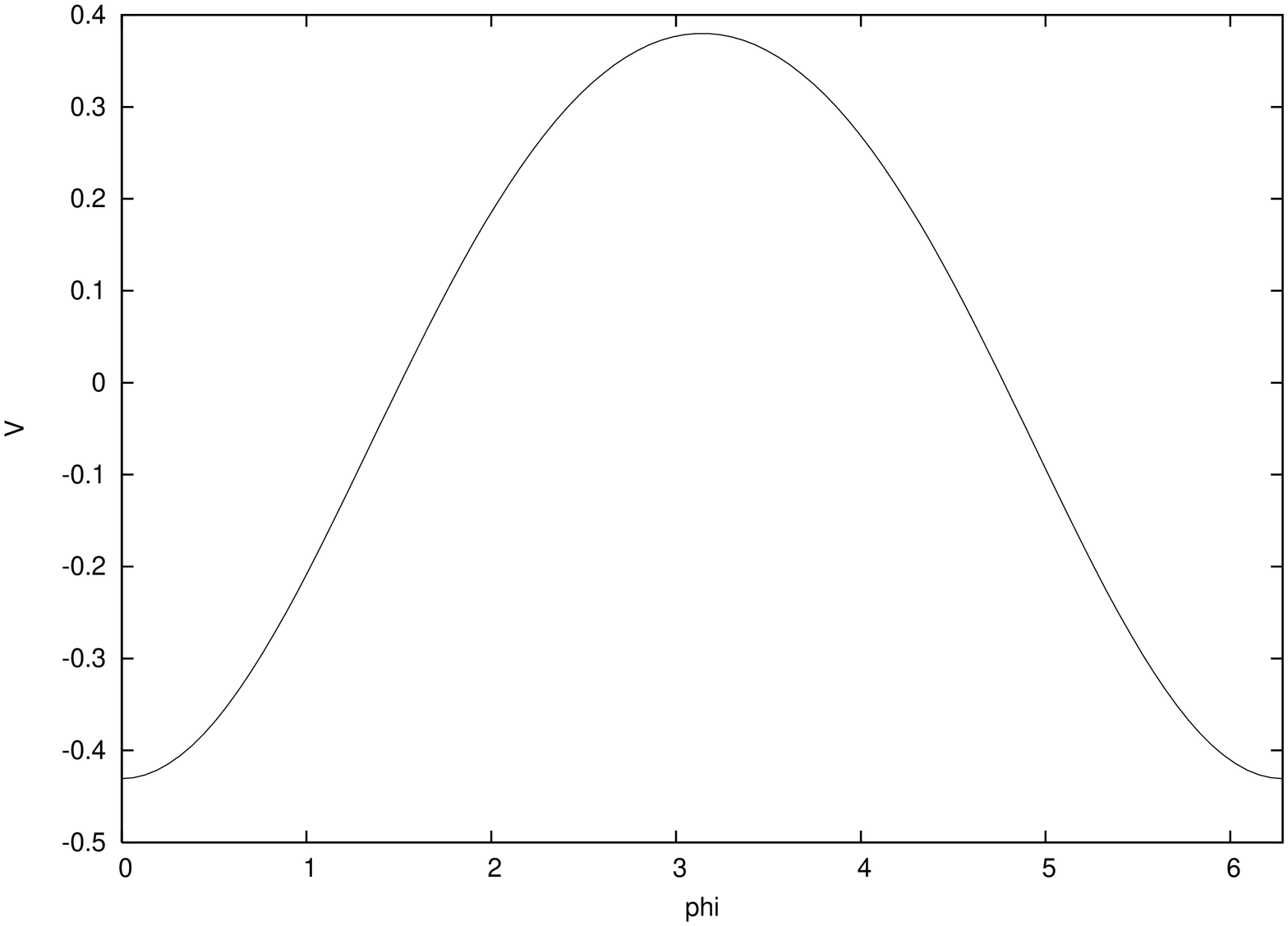}\includegraphics[height=6.8cm,angle=270]{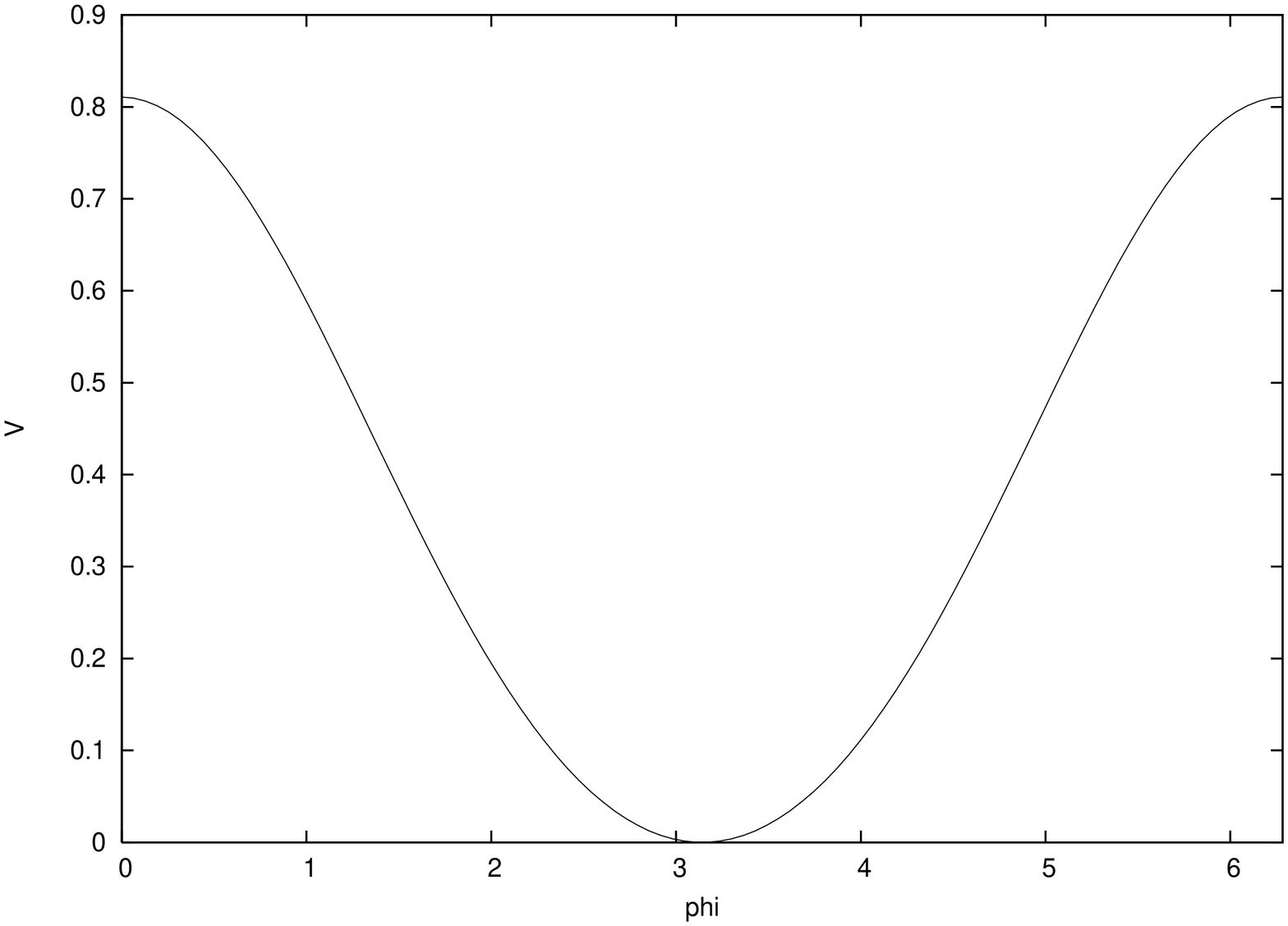}
\caption{\label{fig:pic_weiss} Left panel: perturbative effective
Polyakov-loop potential Eq. \eqref{eq:Weiss_result}
\cite{Weiss:1980rj}. A sketch of a possible form of the effective
potential in the confining phase is given on the right panel.}

}
\end{figure}

A perturbative calculation of the effective potential $V$ for the
order parameter was first performed by Weiss
\cite{Weiss:1980rj}. For this, it is convenient to work with the
``Polyakov gauge'', which rotates the zeroth component $A_0$
of the gauge field into the Cartan subalgebra of SU($N$);
furthermore, the condition $\partial_0 A_0=0$ is imposed.
Focussing solely on the Polyakov loop, it suffices to consider a
single scalar degree of freedom $\phi(\vec x)$, defined by
\be 
A(\vec{x})_\mu ^a=n^a \delta _{\mu 0}\phi(\vec{x}) \label{defphi}.
\ee
Here $n^a$ denotes a constant unit vector in color space,
e.g. $n^a=\delta ^{a3}$. Using this gauge, the order parameter is
fully determined by $\phi$, 
\be \mathcal P(\vec{x}) =
\cos\bigg(\frac{\beta\bar{g}\phi(\vec{x})}{2}\bigg)\, , \label{Pphi}
\ee 
and, consequently, the effective potential can purely be expressed in
terms of $\phi$, being a compact variable $\beta\bar{g}\phi\in[0,2\pi]$. To
leading order in a derivative expansion, the effective potential
yields \cite{Weiss:1980rj}
\be
\label{eq:Weiss_result}
\beta^4 V (\beta\bar{g}\phi) = -\displaystyle \sum_{n=1}^{\infty}
\frac{4}{n^4 \pi^2} \cos(n\beta\bar{g}\phi)  ,
\ee
displayed here as a Fourier-cosine series, and depicted on the left
panel of Fig. \ref{fig:pic_weiss}. The potential has minima at
$\beta\bar{g}\phi=2\pi n$ and is $Z(2)$-symmetric, i.e., invariant
under $\beta\bar{g}\phi\rightarrow2\pi -\beta\bar{g}\phi$. Furthermore
the order parameter is finite for $\beta\bar{g}\phi=2\pi n$ and
therefore the $Z(2)$-symmetry is spontaneously broken and the system
is in the deconfined phase. This perturbative result agrees with the
expectation that perturbation theory holds for high temperature where
the coupling is small. It fails to describe the confinement phase.

In the confined phase, the potential should have its minimum at
$\beta\bar{g}\phi=\pi$, implying a vanishing order parameter. A sketch
of a possible form of the potential with a finite IR regulator (to circumvent, the convexity obstruction
, see below) is shown on the right panel of Fig. \ref{fig:pic_weiss} for illustration.

Various generalizations to Weiss's result have been worked out within
perturbation theory, for instance, the inclusion of a magnetic
background field \cite{star94} or higher-order derivative
expansions \cite{enge98} to name a few. However, in order to investigate the
transition to the confinement phase, reliable nonperturbative tools
are required.  We will base our study on the functional (or ``Exact'')
renormalization group \cite{Wegner} formulated in terms of a flow
equation for the effective action \cite{Wetterich:yh}.

The paper is organized as follows: In Sect. \ref{flow_eq} we briefly review RG
flow equations for Yang-Mills theories and present the flow equation for the
effective potential of the order parameter in a propertime approximation. The
flow of the potential is analyzed in Sect. \ref{results}.  Conclusions and
future directions are discussed in Sect.~\ref{concl}.

\section{Flow Equation for the Polyakov-Loop Potential}\label{flow_eq}

\subsection{Exact Renormalization Group}
In the flow equation approach, we consider the effective average action
$\Gamma _k$ which includes all quantum fluctuations with momenta $|p|>k$, with
the scale $k$ serving as an infrared (IR) regularization. The boundary
condition for the flow equation is fixed at an ultraviolet (UV) scale
$\Lambda$ in terms of the bare action~$\Gamma_{\Lambda}$ to be quantized.
Quantum fluctuations are successively integrated out by lowering the
scale~$k$. In the limit $k\rightarrow 0$, the full quantum effective action
$\Gamma _{k\rightarrow 0}$, i.e., the generating functional of the 1PI Green's
functions, is obtained. The flow of $\Gamma _k$, i.e., the RG trajectory from
the UV scale $\Lambda$ to the deep IR, is obtained from a functional
differential equation~\cite{Wetterich:yh}.

Flow equations for gauge theories require a careful control of gauge
invariance, because the IR regulator scale $k$ introduces sources of
gauge-symmetry breaking in addition to standard gauge-fixing terms.
Nevertheless, standard gauge symmetry can be obtained in the physical limit of
vanishing regulator scale $k\to 0$ by controlling the symmetry constraints
with the aid of regulator-modified Ward-Takahashi identities
\cite{Ellwanger:iz}. In this paper, we employ the flow equation with the
background-field method \cite{Reuter:1993kw,Freire:2000bq} for a simplified
control of gauge invariance within our approximation. We work along the lines
of \cite{Reuter:1993kw,Gies:2002af} where this technique has been used for
zero-temperature gluodynamics. The flow equation for the effective average
action reads \cite{Reuter:1993kw}
\be
\label{eq:flow_eq1}
k\,\partial_{k}\Gamma_{k}[A,\bar{A}]\equiv
\partial_{t}\Gamma_{k}[A,\bar{A}]=\frac{1}{2}\mathrm{Tr}\displaystyle
\frac{\partial_{t}R_{k}(\Gamma_{k}^{ 
    (2)}[\bar{A},\bar{A}])}{\Gamma_{\scriptstyle k}^{ 
    (2)}[A,\bar{A}]+R_{k}(\Gamma_{k}^{ 
    (2)}[\bar{A},\bar{A}])}. 
\ee
Here the trace runs over all internal indices including momenta. The classical
gauge field is given by $A$, whereas the background-field is denoted by
$\bar{A}$ (the ghost fields are not displayed for brevity). $\Gamma_k
^{  (2)}$ denotes the second functional derivative with
respect to the fluctuating fields. The regulator function $R_k$ implements the
IR regularization at the scale $k$, see below. Inserting the background-field
dependent $\Gamma^{(2)}$ into the regulator leads to an adjustment of the
regularization to the spectral flow of the fluctuations as discussed in
\cite{Gies:2002af}, implying a potential improvement when it comes to
approximations.  

The boundary condition for the effective action at the UV scale
$\Lambda$ consists of: 
\be
\Gamma_{\Lambda}[A,\bar{A}]=\Gamma^{cl}[A,\bar{A}]+\Gamma_{\Lambda}^{gf}[A,\bar{A}]
  +\Gamma_{\Lambda}^{gh}[A,\bar{A}], 
\ee 
with the bare Yang-Mills action $\Gamma^{cl}$. The gauge-fixing and ghost
terms are given by
\be
\Gamma_{\Lambda}^{gf}[A,\bar{A}]=\frac{1}{2\xi_{\Lambda}} \int d^{d}
x(D_{\mu}[\bar{A}]a_{\mu})^2,
\quad \Gamma_{\Lambda}^{gh}[A,\bar{A}]=-\!\int d^{d}
  x\bar{c}D_{\mu}[\bar{A}]D_{\mu}[A]c, 
\label{eq:ghofix}
\ee
respectively, with the quantum fluctuations $a_{\mu}$ defined by
$a=A-\bar{A}$. The gauge parameter is denoted by $\xi _k$.

Let us briefly summarize a few properties of the regulator function $R_k$ that
can conveniently be written as
\be\label{eq:reg_def}
R_k(x)=xr(y),\quad y:=\frac{x}{\mathcal{Z}_k k^2}\, ,  
\ee
with $r(y)$ being a dimensionless regulator shape function of dimensionless
argument. Here $\mathcal{Z}_k$ denotes a wave-function renormalization. Note that both
$R_k$ and $\mathcal{Z}_k$ are matrix-valued in field space.

The IR regularization is implemented by the property
\be 
\lim_{x/k^2 \rightarrow 0} R_k(x)=\mathcal Z_k k^2 \quad\Leftrightarrow\quad
r(y)\stackrel{y\rightarrow 0}{\longrightarrow}\frac{1}{y}\,.  
\ee 
The second and third properties of the regulator read 
\be
\lim_{k^2/x \rightarrow 0}
R_k(x)=0\quad\mathrm{and}\quad\lim_{k\rightarrow \Lambda}
R_k(x)\rightarrow\infty \, , 
\ee
and ensure that the regulator is removed in the limit $k\rightarrow 0$
and that the initial bare action is approached at the UV scale $\Lambda$. 
Moreover, the second property guarantees that the regulator vanishes for
modes with $|p|\gg k$, i.e., the theory is not affected by the regulator for
large momenta.

The flow equation (\ref{eq:flow_eq1}) can be mapped onto a generalized
propertime representation, once the background field is identified with the
full quantum field\footnote{This identification itself involves an
  approximation for $\Gamma^{(2)}$ in the denominator of the flow equation, as
  discussed in more detail in \cite{Reuter:1993kw,Litim:2002xm,Gies:2002af}.}
\cite{Litim:2002xm,Gies:2002af}:
\be
\label{eq:fe_PT}
\partial_{t}\Gamma_{k}[A\!=\!\bar{A},\bar{A}]
=\frac{1}{2}\mathrm{STr}\displaystyle \frac{\partial_{t}
  R_{k}(\Gamma_{k}^{  (2)})}{\Gamma_{\scriptstyle
    k}^{  (2)}
  +R_{k}(\Gamma_{k}^{  (2)})}
= \frac{1}{2}\displaystyle \int _0 ^{\infty} ds\,\mathrm{STr} \hat{f} (s,\eta) 
  \exp\big(\frac{s}{k^2}\Gamma_{\scriptstyle k}^{  (2)}\big), 
\ee
where the operator $\hat{f}(s,\eta)$ is given by\footnote{Terms proportional
  to $(\tilde{h}(s)-\tilde{g}(s))$ and $(\tilde{H}(s)-\tilde{G}(s))$ in
  $\hat{f}(s,\eta)$ arise due to the use of the chain rule for $\partial_t
  R_k$ in the flow equation (\ref{eq:flow_eq1}) and manifestly represent terms
  arising from the spectral adjustment of the flow $\sim \partial_t
  \Gamma^{(2)}$.}
\be
\label{eq:fop}
\hat{f}(s,\eta)=\tilde{h}(s)(2-\eta)-(\tilde{h}(s)-\tilde{g}(s))(2-\eta)
  +(\tilde{H}(s)-\tilde{G}(s))\frac{1}{s}\partial _{t}.
\ee
In addition, we have introduced the (matrix-valued) anomalous dimension

\be
        \eta :=-\partial_t \ln \mathcal Z_k =-\frac{1}{\mathcal Z_k}\partial
        _t \mathcal Z_k\, .
\ee
The connection between the function $\hat{f}(s,\eta)$ and the regulator
function $R_k$ is given by the functions $\tilde{g}(s),\,\tilde{G}(s)$ and
$\tilde{H}(s)$ which are defined as:
\be
h(y)=\frac{-yr'(y)}{1+r(y)}\, ,\quad h(y)=\int _0 ^{\infty}
ds\,\tilde{h}(s)\E^{-ys}\, ,\quad\frac{d}{ds}\tilde{H}(s)=\tilde{h}(s)\,
,\quad\tilde{H}(0)=0, 
\ee
\be
g(y)=\frac{r(y)}{1+r(y)}\, ,\quad g(y)=\int _0 ^{\infty}
ds\,\tilde{g}(s)\E^{-ys}\, ,\quad\frac{d}{ds}\tilde{G}(s)=\tilde{g}(s)\,
,\quad\tilde{G}(0)=0\, . 
\ee
The generalized propertime representation (\ref{eq:fe_PT}) of the flow
equation has the advantage that the evaluation of the trace becomes
considerably simplified.

\subsection{Truncated Flow Equations}

Even in the propertime form of Eq. \eqref{eq:fe_PT}, the flow equation cannot
be solved in closed form, which necessitates further approximations. For this,
we truncate the space of action functionals down to a set of operators that
are considered to represent the relevant degrees of freedom for the system or
at least for a particular parametric regime of the system.

For obtaining a first glance at finite-temperature gluodynamics, we
concentrate on the Polyakov-loop potential, employing the simple ansatz:
\be
\label{eq:trunc}
\Gamma_{k}[A,\bar{A}]={\displaystyle \int}d^{d}x\Bigg\{
   \frac{Z_k}{4}F_{\mu\nu}^{a}F_{\mu\nu}^{a}+V_{k}\big((v_{\mu}n^a A_{\mu}
   ^a)^2\big)
 \Bigg\}+\Gamma_{k}^{gf}[A,\bar{A}]+\Gamma_{k}^{gh}[A,\bar{A}]\,.
\ee
Here $v_{\mu}$ denotes the heat-bath velocity, for which we choose
$v_{\mu}=\delta_{\mu 0}$. In the following, we neglect any running in the
ghost- and gauge-fixing sectors, maintaining the form of Eq.
\eqref{eq:ghofix}, $\Gamma^{gf,gh}_k=\Gamma^{gf,gh}_\Lambda$. Furthermore, we
neglect any other gauge-field operators except for the classical action with a
wave function renormalization $Z_k$, and, of course, the Polyakov-loop
potential $V\big((v_{\mu}A_{\mu})^2\big)$.  In order to derive the flow of the
potential, it suffices to evaluate the flow for a trial background field
of the simple form
\be
\label{eq:gauge_choice}
A_\mu ^a=\big(n^a\phi,0\big)^{T}, 
\ee 
where $n^a$ denotes a constant unit vector in color space. Furthermore, we
exploit the freedom to choose suitable wave function renormalizations in the
regulator function for an optimal adjustment of the regulator, cf. Eq.
\eqref{eq:reg_def}, 
\be 
\mathcal Z_k= \left\{
Z_{  k} ^{  gh} =1,\quad
Z^{  L} _{  k}=\frac{1}{\xi}, \quad
Z^{  T} _{  k}\equiv Z_{ 
  k}\right\} 
\ee
for the corresponding ghost, longitudinal and transversal degrees of freedom
with respect to the background field. In particular, we set the transversal
wave-function renormalization equal to the background-field wave-function
renormalization. The choice for $Z^L_k$ renders the truncated flow independent
of the gauge-fixing parameter $\xi$, so that we can implicitly choose the
Landau gauge $\xi _k\equiv0$ which is known to be an RG fixed point
\cite{Ellwanger:1995qf}.  

It is convenient to express the flow equation in dimensionless renormalized
quantities,  
\be
\displaystyle g^2 _{  k}=\displaystyle k^{d-4} Z^{-1}
_{  k} \bar{g} ^2,\quad \displaystyle\varphi=
\beta\bar{g}\phi,\quad \displaystyle v_{  k}= g^2 k^{-d}
V_{  k},   
\ee 
where $\varphi\in[0,2\pi]$. Within our truncation, the flow
equation in $d=4$ dimensions reads \footnote{Details of the
  calculation will be presented in forthcoming publication \cite{GBP}.}
\be
 \partial _{t} v_{  k}(\varphi) 
  =-(4-\eta _{  k})v_{  k}(\varphi)
   +\displaystyle \frac{\alpha _k}{4\pi}  \displaystyle
   \sum_{n=-\infty}^{\infty}
    \Bigg\{ (4-3\eta _{  k})
   \overbrace{\exp\Bigg[-\frac{1}{4}n^{2}\bigg(\frac{k}{T}\bigg)^{2}\Bigg]
     \cos(n\varphi)}^{\mathrm{perturbation\,\,theory}}\nonumber\\ 
  \quad + 4(2-\eta _{  k})\displaystyle 
  \bigg(\frac{T}{k}\bigg){\displaystyle \int_{0}^{\infty}}dxx^{2} 
  \exp\bigg[-\tilde{\omega}_{  n}^{2}-x^{2}
  -\bigg(1-\frac{\tilde{\omega}_{  n} ^{2}}
  {\tilde{\omega}_{ 
 n}^{2}+x^{2}}\bigg)\bigg(\frac{k}{T}\bigg)^2 
  \partial _{\varphi} ^2 {v}_{  k}(\varphi)\bigg] \Bigg\}.
\label{eq:fe_plp}
\ee
Here we have used the abbreviations
\be
\displaystyle\alpha _k=\displaystyle\frac{g_{  k} ^2}{4\pi},
\quad \displaystyle\eta _k=-\displaystyle\partial _t \ln Z _k\, ,\quad 
\displaystyle\tilde{\omega}_{  n}=\displaystyle
2\pi\frac{T}{k} n. 
\ee
Moreover, we have not displayed terms arising from $\propto \partial_t \Gamma^{(2)}$, e.g. terms $\propto \partial _{t} \partial^2_\varphi v_{  k}$, on the right-hand side of the
flow equation. We have dropped these terms in the following preliminary
numerical investigation for reasons of simplicity. As a consequence, the result of
Eq. \eqref{eq:fe_plp} corresponds to a standard propertime flow
\cite{Floreanini:1995aj}. In future work, these terms arising within the Exact
RG flow will be included to facilitate a quantitative study of the differences
between the Exact and the standard propertime RG in the present case. In deriving
Eq. \eqref{eq:fe_plp}, we have furthermore employed a regulator which yields a
particulary simple representation in propertime (or Laplace) space,
\be
        \tilde{h}(s)=\delta (s-1).
\ee
At this point, it is useful to study the overlap of the present result with
perturbation theory. In fact, we rediscover Weiss's result of
Eq. \eqref{eq:Weiss_result} if we (i) hold the coupling fixed,
$\alpha_k=$const., (ii) set the anomalous dimension to zero, $\eta_k=0$, and
(iii) drop the complete second line of Eq. \eqref{eq:fe_plp}. The resulting
simplified equation is an ordinary differential equation, that can immediately
be integrated from $k=\Lambda$ to $k=0$, leading us to the perturbative
one-loop result of Eq. \eqref{eq:Weiss_result}. 

Now, this observation helps estimating the nonperturbative content of the full
Eq. \eqref{eq:fe_plp}: the occurrence of the running gauge coupling $\alpha_k$
and the $k$-dependent anomalous dimension on the right-hand side signal the
``RG improvement'', i.e., a resummation of an infinite set of Feynman
diagrams, performed by the flow equation. Finally, the second line of
Eq. \eqref{eq:fe_plp} depends on (derivatives of) the potential itself. This
term is truly nonperturbative, being induced by fluctuations of the
Polyakov-loop variable on top of its own potential minimum. Since fluctuations
of the degrees of freedom associated with the order parameter become highly
important near the phase transition, we expect that particularly terms of this
kind can cover important aspects of the nonperturbative dynamics.

In principle, the present truncation also allows for a rough calculation of
the running coupling $\displaystyle g _k$ including finite-temperature
effects. For instance, the $\beta$ function for the running coupling using the background field gauge reads
\be
\partial_t g^2_k\equiv \beta_{g^2_k}=\eta_k\, g^2_k,
\ee
and is thus related solely to the background wave function renormalization
that is part of the truncation. However, important features of the running
coupling in the deep infrared require much larger truncations
\cite{Gies:2002af,Pawlowski:2003hq}, hence we decide to take the running coupling as
an external input in this work.

\section{Results}\label{results}
Let us now discuss the flow of the Polyakov-loop potential from the
perturbative UV regime to the IR on the basis of our minimal approximation
given by Eq. (\ref{eq:fe_plp}). In order to solve this partial differential
equation, we rewrite it as an infinite set of coupled first-order differential 
equations by projecting it on a Fourier cosine series; this projection is
naturally suggested by the fact that the $Z(2)$ symmetry of the potential and
its dependence on a compact variable allows for such a Fourier expansion of
the potential itself,
\be
v_k(\varphi)=\sum_{n=0}^\infty v_k^{(n)}\, \cos n\varphi.
\ee
With this procedure, we obtain flow equations for the dimensionless
Fourier coefficients $v _{  k} ^{(n)}$ by
Fourier transforming the right-hand side (RHS) of Eq. \eqref{eq:fe_plp},
\be
\partial _t v_{  k} ^{(n)}=\frac{1}{\pi}\int _0 ^{2\pi} 
d\varphi\,\bigg\{
\text{RHS-Eq.\eqref{eq:fe_plp}}[\varphi,v_k(\varphi)]\bigg\}\,. 
\ee%
For a numerical evaluation of the Polyakov-loop potential, we have to truncate
the infinite set at some $n=N_{max}$ and set all $v_{  k}
^{(n)}$ with $n>N_{max}$ equal to zero. The results, shown in this work, are
calculated for $N_{max}=2$ for simplicity. But we have confirmed that the
inclusion of higher Fourier orders does not change our results
significantly. As the  boundary conditions for the flow equation at the UV
scale $\Lambda=90\,\,\mathrm{GeV}$, we employ the result of 
the one-loop calculation Eq. (\ref{eq:Weiss_result}). \footnote{The integration from
  $\infty$ to $\Lambda$ is controlled by perturbation theory, whereas the
  integration from $\Lambda$ to the deep IR is controlled by the
  nonperturbative flow.}
 
Finally, our simple choice for the running coupling
$g_{  k}$ in this work is
\be
g _{  k} \propto\Big[\ln\big({k}/{\Lambda
  _{c}}\big)\Big]^{-1}, \label{eq:coup}
\ee
where $\Lambda _{c}$ denotes a ``strong scale'' for which we choose $\Lambda
_{c}=1\,\, \mathrm{GeV}$ for simplicity.

\begin{figure}[]
\vspace{-0.5cm}
\centering\includegraphics[width=0.95\linewidth]{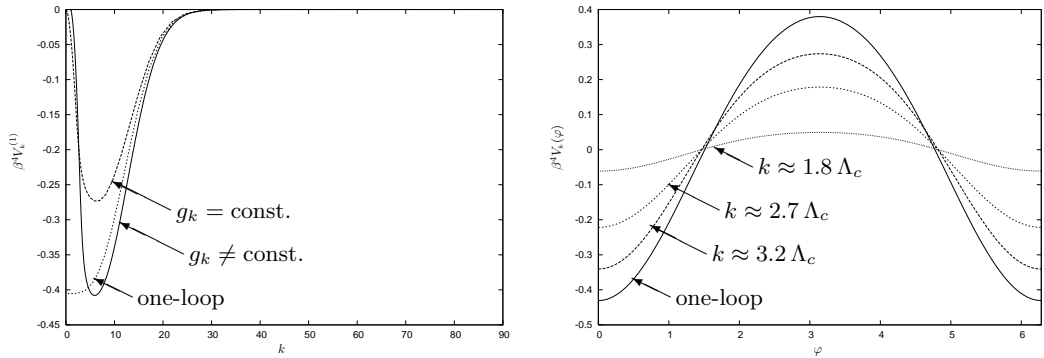}
\caption{\label{fig:pic_res} Left panel: flow of the first Fourier
  coefficient of $V_k =k^{d}g_{  k}
  ^{-2} v_{  k}$ scaled by $T^4$. Right panel: 
  effective potential $V_k$ at different scales $k$ for high temperature
  $T=5\,\Lambda _c$ and $g_{  k}$ as in Eq. \eqref{eq:coup}.}
\end{figure}

In Fig. \ref{fig:pic_res}, we plot the results for the flow of the coefficient
$v _{  k} ^{(1)}$ (left panel) and the Polyakov-loop
potential at different values of $k$ as a function of $\varphi$ for
$T=5~\Lambda _c$ (right panel).  The flow of the first Fourier coefficient
with the running coupling of Eq. \eqref{eq:coup} is compared to that for a
constant coupling $g_k\equiv \mathrm{const.}$ and the one-loop result of Eq.
\eqref{eq:Weiss_result}. We find that the three different calculations are in
good agreement if the scale $k$ is larger than the temperature. In this
regime, the potential is built up by thermal fluctuations. At $k\approx T$, the coefficient of the nonperturbative flow-equation result develops a (negative) minimum and increases to zero for
even smaller $k$. Since we observe a similar behaviour also for all other
Fourier coefficients, the potential flattens for $k<T$,
\be 
v_{  k} ^{(n)}\stackrel{k\rightarrow 0}{\longrightarrow}
0\;(g_k\equiv \mathrm{const.}) \quad\mathrm{and}\quad v _{ 
  k} ^{(n)}\stackrel{k\rightarrow \Lambda _c}{\longrightarrow}
0\;(g_k\,\,\text{of}\,\,\text{Eq. \eqref{eq:coup}}).
\ee
For the running coupling $g_k$, this behavior is depicted on the right panel
of Fig.  \ref{fig:pic_res}. By contrast, the coefficients of the one-loop
result stay finite for $k\rightarrow 0$. This flatness of the potential is, in
fact, nothing but the explicit manifestation of the convexity property that
has to hold for the effective action in general, but is missed by perturbation
theory.

Even though the observation of a convex potential is theoretically highly
satisfactory, it does not tell us anything about the phase of the system. For
this, the resulting expectation value of the Polyakov loop in the IR is
relevant. At this point, we should stress that, even for a finally flat
potential, the expectation value does not remain undetermined, but is
well-defined by its $k\to0^+$ limit (the effective potential is not flat for
any nonzero $k$). However, as we can read off from Fig. \ref{fig:pic_res},  we
do not observe a sign change of the Fourier coefficients under the flow; hence
the minima of the effective potential are $Z(2)$ symmetry breaking and thus
correspond to the deconfinement phase. In the present simple truncation, this
holds true even for lower temperatures. Consequently, our truncation so far is
only capable of describing deconfined dynamics.

\section{Conclusions and Outlook}\label{concl}

We have presented first steps towards a nonperturbative study of the
Polyakov-loop potential based on an RG flow equation. Our intention so far
mainly was to demonstrate the capability of our approach as a matter of
principle by choosing a minimalistic approximation scheme. Already at this
level, we observe important nonperturbative features, such as the
backreactions of order-parameter fluctuations on top of the potential minimum
and convexity of the effective potential. Confronting our first results with
phenomena, we find that our truncation oversimplified the system, since we
find deconfined dynamics on all scales. With hindsight, this is not too
surprising, since, in addition to the Polyakov-loop dynamics, the gluon (and
ghost) sector is basically approximated by its classical form, retaining
``perturbative'' gluon degrees of freedom all the way down to $k=0$.  From
this point of view, our present minimalistic truncation leads to
self-consistent results.

Future work will extend the present approach in various directions: on a
technical level, the influence of the (so far neglected) terms from spectral
adjustment $\propto \partial _t \Gamma _{\scriptstyle k}^{ 
  (2)}$ has to be studied. This is an interesting task by its own, since it
allows a quantitative analysis of the importance of these terms,
distinguishing standard propertime flows from (background-field approximated)
Exact RG flows. On a more conceptual level, the truncation has to be extended to
include a larger class of gluonic (and/or ghost) degrees of freedom, covering
important aspects of nonperturbative dynamics and allowing for an interplay
with the Polyakov-loop sector. For instance, a gluonic potential
$W_k(\frac{1}{4}F_{\mu\nu}F_{\mu\nu})$ replacing the present simple ansatz
$\propto F_{\mu\nu}F_{\mu\nu}$ leaves room for the description of a nontrivial
magnetic sector including gluon condensation as well as complex dynamics in
combination with the Polyakov loop. Future work in this direction is in
progress. 


\begin{theacknowledgments}
  We thank the organizers for the stimulating conference. We are grateful to
  J.M. Pawlowski for useful discussions. J.B. would like to thank the GSI
  Darmstadt for financial support. H.G. acknowledges financial support by the
  Deutsche Forschungsgemeinschaft (DFG) under contract Gi 328/1-2
  (Emmy-Noether program).
\end{theacknowledgments}

\end{document}